\documentclass[12pt]{article}

\usepackage{amsmath}
\usepackage{amssymb}
\usepackage{amsfonts}
\usepackage{lscape}

\begin{document}

\fontsize{12}{6mm}\selectfont
\setlength{\baselineskip}{2em}

$~$\\[.35in]
\newcommand{\dss}{\displaystyle}
\newcommand{\raro}{\rightarrow}

\thispagestyle{empty}

\begin{center}
{\Large\bf Multivortex Solutions of the Weierstrass
Representation}
{\Large\bf }
\end{center}

\begin{center}
{\bf P. Bracken$^{*}$ P. P. Goldstein$^{\dagger}$ and A. M.
Grundland$^{*}$}

$^{*}$ Centre de Recherches Math\'{e}matiques, Universit\'{e}
de Montr\'{e}al,  \\
2920 Chemin de la Tour, Pavillon Andr\'{e} Aisenstadt,   \\
C. P. 6128 Succ. Centre Ville,   \\
Montr\'{e}al, QC, H3C 3J7 Canada   \\

$^{\dagger}$ The Andrzej Soltan Institute for Nuclear Studies,
 \\
Ho$\dot{z}$a 69, 00-681 Warsaw, Poland

e-mail: bracken@CRM.UMontreal.ca, goldstei@fuw.edu.pl,
grundlan@CRM.UMontreal.ca
\end{center}

\vspace{1mm}
\begin{center}
{\bf Abstract}
\end{center}
The connection between the complex Sine and Sinh-Gordon
equations
on the complex plane associated with a Weierstrass type system
and the
possibility of construction of several classes of
multivortex solutions is discussed in detail.
We perform the Painlev\'e test and analyse the possibility
of deriving the B\"acklund transformation from the
singularity analysis of the complex Sine-Gordon equation.
We make use of
the analysis using the known relations for the
Painlev\'{e} equations to construct explicit
formulae in terms of the Umemura polynomials
which are $\tau$-functions for rational solutions
of the third Painlev\'{e} equation. New classes
of multivortex solutions of a Weierstrass system are
obtained through the use of this proposed procedure.
Some physical applications are mentioned in the area of
the vortex Higgs model when the complex Sine-Gordon
equation is reduced to coupled Riccati equations.

\vspace{2mm}
\noindent
PACS subject classifications . Primary 02.30Jr; Secondary
02.30Dk.

\newpage
\noindent
{\bf I. \: INTRODUCTION.}

The complex Sine and Sinh-Gordon equations have been of
considerable
interest recently in many areas of mathematical physics.
They originally appeared in the reduction of the $O(4)$
nonlinear sigma model$^{1,2}$. They have also appeared
in a number of other physical contexts, for example,
in the study of a massless fermion model with chiral
symmetry, and also in the study of the motion of a
vortex filament in an inviscid incompressible fluid$^{3}$.
The equations have been found to be completely
integrable, and some work on the
construction of multi-soliton solutions has been
carried out$^{1-5}$. It has also been shown that the complex
Sine-Gordon theory may be reformulated in terms of
the Wess-Zumino-Witten action and interpreted as the
integrably deformed $SU(2)/U(1)$-coset model, (as in$^{6}$
and references therein).
These studies are based on the complex Sine and Sinh-Gordon
theory in the $(1+1)$-dimensional Minkowski
space-time. These models have
energy functionals which are related to that of the
Ginsburg-Landau energy
$$
E_{GL} = \int [ |\nabla \psi|^{2} + \frac{1}{2}
(1 \pm |\psi|^{2})^{2}] \, d^{2}x.
\eqno(1.1)
$$
>From the physical point of view, this expression
constitutes the basis of the phenomenological
theory of superfluidity$^{7,8}$ and has appeared in
particle physics as well$^{1}$.
The Ginsburg-Landau functional is minimized by the
Gross-Pitaevski vortices$^{7}$. These are topological solitons
of the form
$$
\psi (x,y) = \Phi_{n} (r) e^{in \theta},
\quad \Phi_{n} \rightarrow 1, \qquad r \rightarrow \infty.
$$
It is the purpose here to study vortex solutions
for the equations of interest discussed in this paper.

Another important physical application is to the area
of superconductivity$^{3}$, where vortex solutions play
an essential role. The Lagrangian for the superconducting
system usually takes the form
$$
L= - \frac{1}{4} F_{\mu \nu}^{2} + |\nabla_{\mu} \psi |^{2}
- V(\psi).
\eqno(1.2)
$$
Here, $\psi$ plays the role of the Higgs field,
$F_{\mu \nu}$ pertains to the electrodynamic term
and $V(\psi)$ is the scalar or Higgs potential function
which is responsible for mass generation in the system.

In particular,
the classical Sine-Gordon equation has relevance to
models which are of interest to particle theory.
The complete integrability of the Gross-Neveu model,
in the large $N$ approximation, is quite analogous
to the complete integrability of the classical
Sine-Gordon equation, where $1/N$ plays the role
of the coupling constant$^{1,4}$. For $N=1$, a large coupling
case, the model reduces to the massless Thirring model,
which is scale invariant with anomalous dimensions.
For $N>1$, the theory exhibits non-trivial renormalization
group behavior and mass generation through dimensional
transmutation.

Another area of recent importance with regard to applications
is to the area of liquid crystals and membranes$^{9}$.
Fluid membranes may be idealized as two-dimensional
surfaces in an aqueous solution with each membrane being
made up of a double layer of long molecules.
The curved fluid membrane may be treated as a bending
liquid crystal cell with uniaxial molecular order.
Various physical properties of interest can be calculated
in terms of quantities which are directly related to the
geometry of the surface. For example, in one model for
uniaxial liquid crystals, with the normal to the membrane
$\vec{n}$ of the liquid crystal, it has
been shown$^{9}$ that the elastic energy of curvature
per unit area of the membrane is
$$
F= \int g \, dA,
\qquad
g = \frac{1}{2} k (c_{1} +c_{2} - c_{0})^{2} + \bar{k} c_{1}
c_{2}.
$$
Here, $c_{1}$, $c_{2}$ are the two principal curvatures
of the surface of the membrane, and $c_{0}$ is called
the spontaneous curvature of the surface. The quantity, $F$
is referred to as the total bending energy of the
membrane, the constant $c_{0}$ is related to the asymmetry
of the layers. The positive constant $k$ is the bending
rigidity and $\bar{k}$, which could have either sign,
is the elastic modulus of the Gaussian curvature.
The curvature elastic free energy per
unit area of the membrane can also be formulated rigorously
in terms of two-dimensional differential invariants
of the surface$^{9}$.

This paper is organized as follows.
A generalization of a Weierstrass system
for inducing two-dimensional surfaces in $\mathbb R^{4}$
is presented. It is shown that this system is related
to the complex Sine and Sinh-Gordon equations.
Several properties namely the Lax pair, the Painlev\'e
property and the B\"acklund transformation following
from the singularity analysis, are investigated for the
complex Sine-Gordon equation and then extended to the
Weierstrass
system in Sections 3 and 4. Multivortex
solutions are constructed and the Painlev\'{e} structure
of the associated radial equations is studied. The
$\tau$ functions for the rational class of solutions
of the Painlev\'{e} equation P5 are written in terms
of Umemura polynomials, and explicit forms of such
solutions are given in Section 5.
We also present a particular class of solutions of the
Weierstrass system via the solution of the complex Sine-Gordon
equation.
Section 6 contains examples of solving the Weierstrass
system by means of solutions of the complex Sine-Gordon
equation.

\noindent
{\bf II. \: THE GENERALIZED WEIERSTRASS SYSTEM AND ASSOCIATED
COMPLEX SINE AND SINH-GORDON EQUATIONS.}

\par
The Gauss-Codazzi equations describing a two-dimensional
surface immersed in a three-dimensional sphere which is
itself again immersed into a four-dimensional Euclidean space
has been studied by Darboux$^{10}$. He investigated the
non-linear Dirac-type system for four complex valued functions
$\psi_{i}$ and $\varphi_{i}$, $i=1,2$ satisfying the
following system of equations
$$
\begin{array}{cc}
\partial \psi_{1} = Q_{1} ( \psi_{1} + \dss \frac{\varphi_{1}}
{2 \psi_{2} \bar{\varphi}_{2}}),  &
\bar{\partial} \psi_{2} = Q_{1} \psi_{2},  \\
    \\
\bar{\partial} \varphi_{1} = Q_{2} (\varphi_{1} - \dss
\frac{\psi_{1}}
{2 \varphi_{2} \bar{\psi}_{2}}),  &
\partial \varphi_{2} = Q_{2} \varphi_{2},   \\
    \\
Q_{1} = |\psi_{2}|^{2} \pm |\psi_{1}|^{2}, &
Q_{2} = |\varphi_{2}|^{2} \pm |\varphi_{1}|^{2},   \\
\end{array}
\eqno(2.1)
$$
and its respective complex conjugate equations. The bar
denotes complex conjugate, $\partial= \partial/ \partial z$
and $\bar{\partial} = \partial / \partial \bar{z}$.
The above system can be considered as a variant of the
Weierstrass representation$^{11}$ for surfaces immersed in
$\mathbb R^{4}$ and will refer to it as such.
System (2.1) is a nonlinear first order system of eight
equations, for which eight of sixteen first order
derivatives with respect to $z$ or $\bar{z}$ are known
in terms of functions $\psi_{i}$ and $\varphi_{i}$.
System (2.1) admits several conservation laws such as
$$
\partial (\ln \bar{\psi}_{2}) = \bar{\partial} (\ln \psi_{2}),
\quad
\partial (\ln \varphi_{2}) = \bar{\partial} (\ln
\bar{\varphi}_{2}),
\quad
\partial (\frac{\psi_{1} \bar{\varphi}_{1}}{\bar{\psi}_{2}
\varphi_{2}})
= \bar{\partial} ( \frac{\bar{\psi}_{1} \varphi_{1}}
{\psi_{2} \bar{\varphi}_{2}}).
\eqno(2.2)
$$
As a consequence of conservation laws (2.2), there exist four
real-valued functions $X^{i} (z, \bar{z})$ which are defined
by
$$
\begin{array}{cc}
X^{1} = \int_{\Gamma} \ln \psi_{2} \, dz +
\ln \bar{\psi}_{2} \, d \bar{z},  &
X^{2} = \int_{\Gamma} \ln \varphi_{2} \, dz +
\ln \bar{\varphi}_{2} \, d \bar{z},   \\
   &    \\
X^{3} = \int_{\Gamma} \ln \psi_{2} \varphi_{2} \, dz
+ \ln \bar{\psi}_{2} \bar{\varphi}_{2} \, d \bar{z},    &
X^{4} = \int_{\Gamma} \dss \frac{\bar{\psi}_{1} \varphi_{1}}
{\psi_{2} \bar{\varphi}_{2}} \, dz +
\dss \frac{\psi_{1} \bar{\varphi}_{1}}{\bar{\psi}_{2}
\varphi_{2}} \, d \bar{z},
\end{array}
\eqno(2.3)
$$
for any contour $\Gamma$ in the complex plane
which begins at a fixed $z_{0}$ and ends at variable point
$z$.
The right hand side of (2.3)
does not depend on the choice of the curve $\Gamma$,
since the differential of equations (2.3) are exact ones.
Thus, equations (2.1) and (2.2) allow us to identify the
real-valued functions $X^{i} (z, \bar{z})$, $i=1, \cdots, 4$,
as the coordinates of a surface immersed in
four-dimensional Euclidean space.

In the present paper, we propose a procedure for
constructing explicit multivortex solutions
of system (2.1). These solutions are obtained through
the use of a link between the complex
Sine or Sinh-Gordon equations on the
plane and Weierstrass system (2.1).
To our knowledge, this connection between these two
systems has been observed for the first time here.
At this point, we want to underline that, for the
Weierstrass system (2.1), few explicit solutions
have been found up to now, and this link with
complex Sine-Gordon and complex Sinh-Gordon equations
allows us to construct new classes of solutions explicitly.
We subject system (2.1) to
several transformations in order to simplify its
structure. We start by defining two new
complex valued functions
$$
u = \frac{\psi_{1}}{\bar{\psi}_{2}},
\quad
v = \frac{\varphi_{1}}{\bar{\varphi}_{2}}.
\eqno(2.4)
$$
It is easy to show that if the complex functions
$\psi_{i}$ and $\varphi_{i}$, are solutions of the
first order system (2.1), then the complex-valued functions
$u$
and $v$ defined by (2.4) are solutions of the first order
system of two equations
$$
\partial u = \frac{1}{2} ( 1 \pm |u|^{2}) v,
\qquad
\bar{\partial} v = - \frac{1}{2} (1 \pm |v|^{2}) u,
\eqno(2.5)
$$
and their respective complex conjugate equations.
The elimination of one of the functions $u$ or $v$
in system (2.5) leads to the complex
Sinh-Gordon (CShG) equation when the sign is positive in
(2.5),
and Sine-Gordon (CSG) equation when the sign is
negative in (2.5). Thus we get for both cases
$$
\partial \bar{\partial} u
\mp \frac{\bar{u}}{1 \pm |u|^{2}} \partial u \bar{\partial} u
+ \frac{1}{4} u ( 1 \pm |u|^{2}) = 0.
\eqno(2.6)
$$
As it was shown in$^{1}$, equation (2.6) was derived in the
context of the reduction of the $O(4)$ nonlinear sigma model
and as well, the reduction of the self-dual Yang-Mills
equations
and relativistic equations$^{2,12,13}$.

Note that if $v$ tends to one in CSG equations (2.5),
then $Q_{2}$ vanishes and
system (2.1) takes the form
$$
\partial \psi_{1} = Q_{1} ( \psi_{1} + \frac{1}{2 \psi_{2}}),
\quad
\bar{\partial} \psi_{2} = Q_{1} \psi_{2}.
\eqno(2.7)
$$
Conversely, if $u$ tends to one in CSG equations (2.5),
then $Q_{1}$ vanishes and system (2.1) becomes
$$
\bar{\partial} \varphi_{1} = Q_{2} (\varphi_{1} - \frac{1}{2
\varphi_{2}}),
\quad
\partial \varphi_{2} = Q_{2} \varphi_{2}.
\eqno(2.8)
$$
These limits characterize the properties of the solutions of
system (2.1).

Now, we discuss a set of conditions which allow the system
(2.1) to become a decoupled system of equations.

{\bf Proposition 1.} Let the complex functions $u$ and $v$
be solutions of system (2.5). Let the functions
$\psi_{i}$ and $\varphi_{i}$ be defined in terms of
$u$ and $v$ by
$$
\begin{array}{cc}
\psi_{1} = \epsilon u ( 1 \pm |u|^{2})^{-1/2}  &
\varphi_{1} = \epsilon v (1 \pm |v|^{2})^{-1/2},   \\
       &        \\
\psi_{2} = \epsilon (1 \pm |u|^{2})^{-1/2},
&
\varphi_{2} = \epsilon ( 1 \pm |v|^{2})^{-1/2},
\quad \epsilon = \pm 1.
\end{array}
\eqno(2.9)
$$
Then the general integrals of system (2.1) are given by
$$
\begin{array}{ccc}
\psi_{1} = u \bar{A} (\bar{z}) e^{z},  &  & \varphi_{1} = v
B(z) e^{\bar{z}}, \\
      &      &        \\
\psi_{2} = A(z) e^{\bar{z}},    &    &  \varphi_{2} = \bar{B}
(\bar{z}) e^{z}, \\
\end{array}
\eqno(2.10)
$$
where the complex functions $A$ and $B$ satisfy the following
conditions,
$$
|A|^{2} = e^{-(z + \bar{z})} (1 \pm |u|^{2})^{-1},
\qquad
|B|^{2} = e^{-(z + \bar{z})} (1 \pm |v|^{2})^{-1}.
\eqno(2.11)
$$

{\bf Proof:} Substituting (2.4) into system (2.1), we obtain
an overdetermined system for the functions $\psi_{2}$ and
$\varphi_{2}$ of the following form
$$
\begin{array}{ccc}
\partial (u \bar{\psi}_{2}) = (1 \pm |u|^{2}) |\psi_{2}|^{2}
(u \bar{\psi}_{2} + \dss \frac{v}{2 \psi_{2}}),  &  &
\bar{\partial} \psi_{2} = (1 \pm |u|^{2}) |\psi_{2}|^{2}
\psi_{2},  \\
      &          &         \\
\bar{\partial} (v \varphi_{2}) = (1 \pm |v|^{2})
|\varphi_{2}|^{2}
(v \bar{\varphi}_{2} - \dss \frac{u}{2 \varphi_{2}}),  &   &
\partial \varphi_{2} = (1 \pm |v|^{2}) |\varphi_{2}|^{2}
\varphi_{2}.  \\
\end{array}
\eqno(2.12)
$$
Consider the first equation in the first line of (2.12).
Expanding the derivative term $\partial (u \bar{\psi}_{2})$
and using (2.5), this equation reduces to the form
$$
\partial \bar{\psi}_{2} = ( 1 \pm |u|^{2}) |\psi_{2}|^{2}
\bar{\psi}_{2}.
$$
Using (2.9) to replace $|\psi_{2}|^{2}$ in this result, it
reduces to
$\partial \bar{\psi}_{2} = \psi_{2}$. Making use of (2.9), the
remaining three equations in (2.12) can be treated in a
similar way.
Thus the initial system (2.1) becomes a linear system of the
form
$$
\begin{array}{ccc}
\bar{\partial} \psi_{2} = \psi_{2},  &   &  \partial
\varphi_{2} = \varphi_{2}. \\
\end{array}
\eqno(2.13)
$$
These two equations can be easily integrated to give
$\psi_{2}$ and
$\varphi_{2}$ as given in (2.10), then (2.4) can be used to
obtain
$\psi_{1}$ and $\varphi_{1}$. The results in (2.10) must be
consistent with those in (2.9). If we calculate the modulus of
$\psi_{2}$ and $\varphi_{2}$ from (2.10) and equate to the
modulus calculated from (2.9), the conditions (2.11) are
obtained.
In fact, equating $\psi_{i}$, $\varphi_{i}$ in (2.10) to
their corresponding forms in (2.9), we must also have that
$$
\bar{A}(\bar{z}) e^{z} = \epsilon (1 \pm |u|^{2})^{-1/2}=A(z)
e^{\bar{z}},
\qquad
B(z) e^{\bar{z}} = \epsilon (1 \pm |v|^{2})^{-1/2}
= \bar{B} ( \bar{z}) e^{z}.
$$
A set of differential constraints which must be satisfied
can be obtained by substituting (2.9) into (2.13), and we find
$$
(\bar{\partial} u \bar{u} + \frac{1}{2} uv ( 1 \pm |u|^{2}))
= \mp 2 (1 \pm |u|^{2}),
\qquad
(\partial v \bar{v} - \frac{1}{2} \bar{u} v ( 1 \pm |v|^{2}))
=
\mp 2 (1 \pm |v|^{2}).
$$
QED

>From the computational point of view, it is more convenient
to deal with the CShG or CSG equations (2.5) than with the original
system (2.1). From every solution of CShG or CSG equations
(2.5), we
can integrate a linear system (2.13), and consequently, a
very large class of solutions of system (2.1) can be found
explicitly by making use of formulae (2.10) and (2.11).

Using the connection between the CSG equation (2.6)
and Weierstrass system (2.1), we discuss in the next section
in detail the
Painlev\'e analysis of the CSG equation which
allows us to extend this analysis to the Weierstrass system.

{\bf III. PAINLEV\'{E} ANALYSIS OF THE COMPLEX SINE-GORDON
EQUATION.}

Integrability of the CSG equation (2.6) is confirmed by tests for
the Painlev\'e property. We perform the classical test of$^{14}$
extended to partial differential equations in$^{15}$, assuming a
solution in the form of a Laurent series about an arbitrary
singularity manifold $F( z, \bar{z}) =0$ and checking
compatibility of the resulting recurrence formulae. Detailed
discussion of the meaning and validity of this test may be found
in$^{20}$. The test is carried out for the CSG equation (2.6) and
for the system (2.5). Both versions of the CSG pass the test. For
the system (2.5), we discuss the possibilities and limitations of
the approach to the B\"{a}cklund transformation$^{15}$.

Both those forms of the CSG contain non-analytic expressions,
therefore $z$ and $\bar{z}$ should be treated as two
independent variables and extended to two separate complex
planes. Similarly, the functions $u$ and $\bar{u}$ need not be complex
conjugates of each other, when their independent
variables are separately extended. We assume 
also the same holds for $v$ and $\bar{v}$. To avoid the misleading
complex conjugate symbol, we denote $\bar{z}$ by $t$,
$\bar{u}$ by $w$ and $\bar{v}$ by $s$, while symbols of the
derivatives $\partial$ and $\bar{\partial}$ will be replaced
by alphabetic subscripts $z$ and $t$, respectively.

In this notation, the original CSG  equation (2.6), is cast
into polynomial form. It has the form of the system
$$
\begin{array}{c}
(1 - uw) u_{zt} + w u_{z} u_{t} + \frac{1}{4} u ( 1 - u w)^2 =
0,  \\
     \\
(1 - uw) w_{zt} + u w_{z} w_{t} + \frac{1}{4} w ( 1 - uw)^2 =
0.
\end{array}
\eqno(3.1)
$$

Both dependent variables, $u$ and $w$, begin their series with
$F$ inverse. The coefficients at this power of $F$ may be
written as
$$
u_{0} = 2 (F_{z} F_{t})^{1/2} / Q_{0},
\quad
w_{0} = 2 (F_{z} F_{t})^{1/2}  Q_{0},
\eqno(3.2)
$$
where $Q_{0}$ is an arbitrary function of $z$ and $t$.

The indices, at which the determinant of the linear recurrence formulae turns
zero,
are $-1$ (corresponding to arbitrariness of $F$), $0$ (due to arbitrariness of
$Q_{0}$), $1$ and $2$. Both compatibility conditions at the last two indices
are
satisfied. As the $2 \times 2$ matrices of the respective linear systems have
rank one
at each of these indices, each of the recurrence formulae yields one arbitrary
function (say, $Q_{1}$ and $Q_{2}$), to complete the set of four first
integrals $(F,
Q_{0}, Q_{1}, Q_{2})$ in the general solution of (2.6). To conclude that the
CSG
equation (2.6) has the Painlev\'e property, we should also examine the
singularity $u
w \to 1$. This task is straightforward when working with system (2.5).

The system (2.5) is just a form of the CSG equation (2.6),
so it should pass the same test. We describe it in more detail
because of the close connection of (2.5) with the Weierstrass system
(2.1).

For the purpose of the ``Painlev\'e test'', equation (2.5) is
a $4 \times 4$ system (we choose the minus sign for
further analysis) given by
$$
\begin{array}{cc}
u_{z} - (1 - wu)v/2 =0, &
w_{t} - (1 - wu)s/2 =0,  \\
  \\
v_{t} + (1 -sv)u/2 = 0, &
s_{z} + (1 - sv) w/2 =0.
\end{array}
\eqno(3.3)
$$
We may choose $-1$ as the initial exponents in the
Laurent series of either $u$ and $w$ or $v$ and $s$.
We perform the calculations for the former choice, which
corresponds to our previous analysis of the
CSG equation (2.6); the latter choice is equivalent to it by
exchange of $u$ with $v$, $w$ with $s$ and $z$ with $-t$. The
above-mentioned singularity $u w \to 1$ corresponds to the
latter choice.
The resulting initial exponents of $v$ and $s$ are both equal
to $0$ and their Laurent expansions turn into the
Taylor ones. Coefficients $u_{0}$ and $w_{0}$ are
the same as those in (3.2), while
$$
v_{0} = (F_{z}/ F_{t})^{1/2} / Q_{0},
\quad
s_{0} = (F_{t}/F_{z})^{1/2} Q_{0}.
\eqno(3.4)
$$

At this point, we would like to mention the following fact. Usually, the classical
``Painlev\'e test'' is not possible when terms of lowest order in $F$ have nonnegative
exponents. Such terms may lack the property (necessary for the algorithm of$^{14}$
and$^{15}$) that differentiation decreases their order of magnitude by one. However,
here the derivatives of $v$ and $s$ do not contribute to the dominant terms in the
last two equations of (3.3) and that deficiency is meaningless.

The remaining terms are derived from the recurrence formulae
$$
\begin{array}{c}
n F_{z} u_{n} + (F_{z}/ Q_{0}^2) w_{n} + 2 F_{z} F_{t} v_{n}
= - (u_{n-1})_{z} + \frac{1}{2} v_{n-2}  \\
-\frac{1}{2} u_{0} \sum_{k=1}^{n-1} w_{k} v_{n-k}
-\frac{1}{2} \sum_{k=0}^{n-1} \sum_{l=1}^{n-k-1}
w_{k} u_{l} v_{n-k-l},    \\
\end{array}
\eqno(3.5 a)
$$
$$
\begin{array}{c}
(F_{t} Q_{0}^{2}) u_{n} + n F_{t} w_{n} + 2 F_{z} F_{t} s_{n}
= - (w_{n-1})_{t} + \frac{1}{2} s_{n-2}  \\
- \frac{1}{2} w_{0} \sum_{k=1}^{n-1} u_{k} s_{n-k}
- \frac{1}{2} \sum_{k=0}^{n-1} \sum_{l=1}^{n-k-1}
u_{k} w_{l} s_{n-k-l},
\end{array}
\eqno(3.5 b)
$$
$$
\begin{array}{c}
(n-1) F_{t} v_{n} - (F_{z}/Q_{0}^{2}) s_{n} = - (v_{n-1})_{t}
\\
+ \frac{1}{2} u_{0} \sum_{k=1}^{n-1} s_{k} v_{n-k}
+ \frac{1}{2} \sum_{k=0}^{n-1} \sum_{l=1}^{n-k-1} s_{k} u_{l}
v_{n-k-l},
\end{array}
\eqno(3.5 c)
$$
$$
\begin{array}{c}
- (F_{t} Q_{0}^{2}) v_{n} + (n-1) F_{z} s_{n} = -
(s_{n-1})_{z}  \\
+ \frac{1}{2} w_{0} \sum_{k=1}^{n-1} v_{k} s_{n-k}
+ \frac{1}{2} \sum_{k=0}^{n-1} \sum_{l=1}^{n-k-1} v_{k} w_{l}
s_{n-k-l},
\end{array}
\eqno(3.5 d)
$$
where $v$ and $s$ with negative subscripts are both set equal
to
zero, and have been admitted for compact notation.

The determinant of the system (3.5) reads
$$
(F_{z} F_{t})^{2} (n+1) n (n-1) (n-2),
\eqno(3.6)
$$
Hence, the indices at which it becomes zero, are $-1$, $0$,
$1$ and $2$ as in the previous case of the CSG equation (2.6).
Similarly, all the compatibility conditions are satisfied,
whence we conclude that equation (2.5) has the Painlev\'e
property.

This Painlev\'e integrability obviously extends to the
Weierstrass system (2.1) as the Painlev\'e property is
invariant
under the homographic transformation which converts the
equations (2.4) to (2.1).

The method of the Laurent expansion may often be extended to deriving the B\"{a}cklund
transformation and further an explicit integration scheme$^{15}$. The usual approach
relies on truncation of the Laurent series at the term of order $F^{0}$, which is
assumed to satisfy the original equation. The truncation usually is possible through
appropriate choice of the arbitrary functions (first integrals). Some extra
assumptions on the coefficients and expansion variable may also be necessary. A
systematic approach to that problem may be found in $^{16,20}$.

However, the usual method contains assumptions which are too
restrictive
for application to equations (3.3). Namely, the Laurent series
of $v$ and $s$, which begin with the $F^{0}$ terms, reduce to
a single term each. This means that $v$ and $s$ would not be
transformed at all. Moreover, the truncation at $F^{0}$
implies vanishing
of terms proportional to $F^{1}$. This imposes further
constraints
on these variables: from their recurrence equations (3.5) at
$n=1$
$$
- (F_{z}/Q_{0}^{2}) s_{1} = - (v_{0})_{t} = 0,
\eqno(3.7 a)
$$
$$
- (F_{t} Q_{0}^{2}) v_{1} = - (s_{0})_{z} = 0.
\eqno(3.7 b)
$$
It follows that $v_{0}$ should be independent of $t$, while
$s_{0}$
should be independent of $z$. However, these coefficients are
reciprocals of each other (3.4), whence neither of them may
depend on $z$ or $t$. This, together with the truncation of the
series, reduces $u$ and $v$ to constants. If we denote
$$
s=k,  \qquad  v= 1/k,  \qquad  k=constant,
\eqno(3.8)
$$
then the original equations (3.3) reduce to a system
of coupled Riccati equations
$$
u_{z} - ( 1 - wu)/ (2k) =0,  \qquad
w_{t} - ( 1 - wu)k/2 = 0,
\eqno(3.9)
$$
which may immediately be linearized by substitution
$$
u = (2/k) (\ln \Psi)_{t}, \qquad
w= 2k (\ln \Psi)_{z},
\eqno(3.10)
$$
to the Helmholtz equation
$$
\Psi_{zt} = (1/4) \Psi.
\eqno(3.11)
$$
Obviously, this linearization also yields an (almost trivial)
transformation of $u$ and $v$, a superposition principle, and
other properties.

 The method of the Laurent expansion starting
from eqs. (3.1) is free of those shortcomings. However the condition of vanishing of
the higher order coefficients results in cumbersome equations, which makes the
procedure hardly worthwhile in comparison with other methods of integration (see next
section).

{\bf IV. ON EQUIVALENCE OF TWO FORMS OF THE COMPLEX
SINE-GORDON EQUATION.}

Throughout this paper we investigate the CSG equation in the
form
of equations (2.6).
Another form of the CSG equation was given in$^{17,18}$ as
follows
$$
(\frac{q_{\xi}}{\sqrt{1 + qp}})_{\eta} = 4 q,
\qquad
(\frac{p_{\xi}}{\sqrt{1+ qp}})_{\eta} = 4 p.
\eqno(4.1)
$$
Let
$$
u = \sin (\Phi/2) \exp(i \alpha).
\eqno(4.2)
$$
The polar coordinates $\Phi$ and $\alpha$ satisfy a system of
equations
similar to that given by Lund$^{4}$
$$
\partial \bar{\partial} \Phi
- 2 \frac{\sin(\Phi/2)}{\cos^3 (\Phi/2)} \partial \alpha
\bar{\partial} \alpha + \frac{1}{4} \sin \Phi =0,
\eqno(4.3)
$$
$$
\partial (\tan^2 (\Phi/2) \bar{\partial} \alpha)
+ \bar{\partial} ( \tan^2 (\Phi/2) \partial \alpha) = 0.
\eqno(4.4)
$$

We formulate the following statement for systems (2.6) and
(4.1).

{\bf Proposition 2.} Equation (4.1) is transformed into CSG
equation
(2.6) through the following relations
$$
\xi = \frac{1}{4} \bar{z} ,   \quad
\eta = - \frac{1}{4} z,       \quad
q = - \sin \Phi \, \exp(-i \beta),  \quad
p = \sin \Phi \exp( i \beta),
\eqno(4.5)
$$
where the phase is given by
$$
\beta = \int^{z}_{z_{0}} \{ [ 1 + \tan^2 (\Phi(z',
\bar{z})/2)]
\partial' \alpha (z', \bar{z}) \, d z'
+ [ 1 - \tan^2 (\Phi (z, \bar{z}')/2)] \bar{\partial}'
\alpha (z, \bar{z}') \, d \bar{z}' \}.
\eqno(4.6)
$$
The lower limit of integration, $z_{0}$,
is fixed and depends on the initial conditions.

{\bf Proof.} Substitution of (4.5) and (4.6) into (4.1) yields
the system (4.3) and (4.4).

Note that both equations, (2.6) and (4.1) depend on their
phases
$\alpha$ and $\beta$ respectively, through their derivatives
only,
except for linear dependence on factors $\exp( i \alpha)$ and
$\exp( i \beta)$. Therefore, any change of $\beta$, which
leaves
its derivatives unchanged, for example, one that arises from
deformation of the integration contour in (4.6), does not
affect
equivalence of those equations. Moreover, for those $u$,
which satisfy the CSG equation (2.6), the integrand is an exact
differential and the path of integration does not even affect
the value of the phase.

Note also that the form of the equation determining evolution
of the phase
(4.4) suggests integration in terms of an arbitrary potential
$\psi (z, \bar{z})$
$$
\tan^2 (\Phi/2) \partial \alpha = \partial \psi,
\qquad
\tan^2 (\Phi/2) \bar{\partial} \alpha = - \bar{\partial} \psi.
\eqno(4.7)
$$
However, the potential is not arbitrary
since the compatibility condition
$\partial \bar{\partial} \alpha = \bar{\partial} \partial
\alpha$
imposes a constraint of the form similar to the original phase
equation (4.4)
$$
\partial (\cot^2 (\Phi/2) \bar{\partial} \psi)
+ \bar{\partial} ( \cot^2 (\Phi/2) \partial \psi ) = 0.
\eqno(4.8)
$$
Obviously, if $\alpha$ is a solution of (4.4) for a given
$\Phi$
then $\psi$ solves the same equation for $\Phi$ shifted by an
odd
multiple of $\pi$ or subtracted from such a multiple.
However, this is not a symmetry of the CSG equation (2.6), as
the
signs in the amplitude equation (4.3) are changed by such a
transformation. Finally, repetition of the transformation
brings us back to the original equation. A similar property
holds for the CSG equation in the form (4.1), for which
the phase equation may be written as
$$
\partial ( \cos^{-1} \Phi \bar{\partial} \beta)
- \bar{\partial} ( \cos \Phi \partial \beta) = 0,
\eqno(4.9)
$$
where the change of independent variables has already been
performed.

This symmetry has an additional consequence. The
transformations (4.5) and (4.6)
from equations (2.6) to (4.1) have some features of a
B\"acklund
transformation. Namely, the definition of $\beta$ by means of the contour
integral (4.6) is a solution of a coupled pair of
differential equations
$$
\partial \beta = [1 + \tan^2 (\Phi/2) ] \partial \alpha,
\qquad
\bar{\partial} \beta = [1 - \tan^2 (\Phi/2)] \bar{\partial}
\alpha.
\eqno(4.10)
$$
System (4.10) is overdetermined, as the right hand sides of
the
above equations must satisfy the compatibility condition
$\partial \bar{\partial} \beta = \bar{\partial} \partial
\beta$.
This condition is indeed satisfied as it proves to be
equivalent
to equation (4.4). Thus we obtain (4.4) in two ways: either
through direct substitution of (4.10) to (4.9), or from
the above compatibility condition.

The transformation defined by (4.5) and (4.6) may be extended
to the inverse
scattering method. The inverse scattering technique for
(4.1) was given in$^{17}$. Using (4.5) and (4.6), we
obtain the Lax pair for (2.6)
$$
\partial X = - \frac{1}{4 \lambda} [ Y_{1}, X],  \qquad
\bar{\partial} X = \frac{1}{4} [Y_{2} + \lambda Y, X],
\eqno(4.11)
$$
where
$$
Y= \left(  \begin{array}{cc}
-i  &   0   \\
0   &   i   \\
\end{array}   \right),  \quad
Y_{1} =  \left(   \begin{array}{cc}
i \cos \Phi   &   \sin \Phi  e^{i \beta}   \\
- \sin \Phi e^{- i \beta} & - i \cos \Phi  \\
\end{array}   \right),
$$
$$
Y_{2} = 2i \left(    \begin{array}{cc}
0   &  \dss \frac{\bar{\partial}(\sin \Phi e^{i \beta})}{\cos
\Phi}  \\
\dss \frac{\bar{\partial}(\sin \Phi e^{-i \beta})}{\cos \Phi}
&   0  \\
\end{array}   \right).
\eqno(4.12)
$$
Further extension to the Weierstrass system (2.1) is also
possible, by means of the transformation (2.4). The matrices
$Y_{1}$ and $Y_{2}$ expressed in terms of $u$ become
$$
Y_{1} = \left(   \begin{array}{cc}
i (1 - 2 u^2 e^{-2i \alpha})  &  \sqrt{1- (1 - 2 u^2 e^{-2 i
\alpha})^2}
e^{i \beta}   \\
     &        \\
- \sqrt{1- (1 - 2 u^2 e^{-2 i \alpha})^2} e^{-i \beta} &
-i (1- 2 u^2 e^{-2 i \alpha})  \\
\end{array}   \right)
$$
$$
Y_{2} = \left(   \begin{array}{cc}
0   &  \dss \frac{\bar{\partial}(\sqrt{1- (1- 2 u^2 e^{-2i
\alpha})^2} e^{i \beta})}
{1- 2 u^2 e^{-2 i \alpha}}   \\
\dss \frac{\bar{\partial}(\sqrt{1- (1- 2 u^2 e^{-2i
\alpha})^2} e^{-i \beta})}
{1 - 2 u^2 e^{-2 i \alpha}}  &   0   \\
\end{array}   \right).
\eqno(4.13)
$$
>From (2.4), complex functions $u$ and $v$ are given in terms
of $\psi_{1}$, $\bar{\psi}_{2}$ and $\varphi_{1}$,
$\bar{\varphi}_{2}$,
respectively. Taking into account that functions $u$, $v$
satisfy the
same equation (2.6), the resulting Lax pair (4.11) for
Weierstrass
system (2.1) can be described by a system of five two by two
matrices
$Y_{1}$, $Y_{2}$ in terms of $(\psi_{1},\bar{\psi}_{2})$ and
$(\varphi_{1},
\bar{\varphi}_{2})$ respectively, and the constant matrix $Y$.

{\bf V. \: MULTIVORTEX SOLUTIONS}

\par
At this point, we would like to derive, through the link
between first order system (2.1) and equations (2.6), a
procedure for
constructing multivortex solutions in explicit form.
Now, we concentrate on certain class of multivortex
solutions of equations (2.6) in polar coordinates
$(r, \theta)$ on the plane determined by
$$
u = A_{n} (r) e^{i n \theta},  \quad
n \in \mathbb Z.
\eqno(5.1)
$$
Equation (2.6), under the assumption (5.1) is reducible
to a second order ODE of the form
$$
\frac{d^{2} A_{n}}{d r^{2}}
+ \frac{1}{r} \, \frac{d A_{n}}{dr} \mp
\frac{A_{n}}{1 \pm A_{n}^{2}}
[ (\frac{d A_{n}}{dr})^{2} \pm \frac{n^{2}}{r^{2}}]
+ (1 \pm  A_{n}^{2}) A_{n} = 0.
\eqno(5.2)
$$
By a homographic transformation of the dependent variable
$$
A_{n} = c \frac{1 + w(z)}{1- w(z)},
\quad
z= r,
\eqno(5.3)
$$
where $c=-i$ for the CShG system, and $c=1$ for the CSG
system, respectively,
equation (5.2) has the structure of the fifth Painlev\'{e}
(P5)
equation
$$
w'' = \frac{3w -1}{2w (w-1)} w^{'2} - \frac{w'}{z}
+ \frac{(w-1)^{2}}{z} ( \alpha w + \frac{\beta}{w})
+ \frac{\gamma}{z} w + \delta
\frac{w (w+1)}{w-1},
\eqno(5.4)
$$
with the coefficients $\alpha$ and $\beta$ parametrized by a
number $n \in \mathbb Z$ and $\gamma, \delta$ fixed as follows
$$
\alpha = - \beta = \frac{n^{2}}{8},
\quad
\gamma = 0,
\quad
\delta =-2.
\eqno(5.5)
$$
Such reduction to P5 has recently been performed$^{19}$.
In general, equation P5 is not integrable in terms of
known classical transcendental functions. However, for
specific values of the parameters, solutions of equation (5.4)
can be
reduced to two types of nontranscendental functions,
that is, to solutions of a Riccati equation with one
arbitrary parameter or to three types of rational solutions
of equation P5 $^{20,21}$. According to$^{21}$,
equation (5.4) with coefficients (5.5) can be written in
equivalent form as
a first order system of ODEs
$$
\begin{array}{c}
z \dss \frac{d p }{d z} = - \frac{\epsilon n}{2} - \epsilon n
p
- p q - p^{2} q,   \quad   \epsilon = \pm 1,   \\
   \\
z \dss \frac{d q}{d z} = -2 z^{2} + \epsilon n q - 4 z^{2} p
+ \frac{q^{2}}{2} + p q^{2},   \\
\end{array}
\eqno(5.6)
$$
where $p = w/(1-w)$. The function $q(z)$ satisfies a
Painlev\'{e} type equation of the form
$$
q'' = \frac{q}{q^{2} -4 z^{2}} q^{'2}
- \frac{q^{2} +4 z^{2}}{q^{2} - 4 z^{2}} \cdot
\frac{q'}{z} +
\frac{q}{4 z^{2} (q^{2} - 4 z^{2})}
[16 n z^{2} ( 2 \epsilon - n) - (q^{2} - 4 z^{2})^{2}].
$$
The function $q^{2} - 4 z^{2}$ has two roots at $q = 2z$.
Using the transformation
$$
y(z) = \frac{q + 2z}{q-2z},
\quad
q \neq 2z,
$$
we get that $y(z)$ is also a solution of equation P5 with
parameters
$$
\tilde{\alpha} = - \tilde{\beta}=
\frac{(1- \epsilon n)^{2}}{8},
\quad
\tilde{\gamma} = 0,
\quad
\tilde{\delta} =-2.
\eqno(5.7)
$$
Propositions 3 to 5 are special cases studied by V.
Gromak$^{21}$
(Chapter 12, section 14) concerning the fifth Painlev\'{e}
equations
with specific parameters. This analysis is
used to construct solutions to Weierstrass system (2.1).

{\bf Proposition 3}. Let $w= w(z)$ be a solution of the fifth
Painlev\'e
equation P5 (5.4)
with parameters given by (5.5), such that the function
$$
\Phi_{1} (w) \equiv z w' - \frac{\epsilon n}{2} w^{2}
+ 2 z w + \frac{\epsilon n}{2} \neq 0,
\eqno(5.8)
$$
does not vanish for any $n \in \mathbb Z$,
then the function
$$
w_{1} = 1 - \frac{4z}{\Phi_{1} (w)},
\eqno(5.9)
$$
is a solution of the fifth Painlev\'e
equation (5.4) with parameters given by (5.7).

Proposition 3 establishes the Auto-B\"{a}cklund transformation
(Auto-BT)
for equation P5 when $\gamma=0$, $\delta = -2$ and $\alpha = -
\beta$
are parametrized by $n \in \mathbb Z$.

Now, let us discuss the link
between equations P5 with different values of the
parameter $\delta$, namely, $\delta \neq 0$ and $\delta = 0$.
Note that for $\delta \neq 0$, the solutions of equation (5.4)
are expressible in terms of Bessel functions, whereas for
$\delta =0$, they can be expressed in terms of Umemura
polynomials, which is presented below.

{\bf Proposition 4}. Let $u(z) \neq 0$ be a solution of
equation P5
with parameters given by (5.5). Then the function
$$
\tilde{u} (z) = \frac{f^{2} (\sqrt{z})}
{f^{2}( \sqrt{z}) - 1},
\eqno(5.10)
$$
where $f(z)$ is defined by
$$
f(z) = \frac{d}{dz} \ln u(z) - \frac{n}{4z} ( u(z) -
\frac{1}{u(z)}),
\quad
n \in \mathbb Z,
$$
is a solution of equation P5 with parameters
$$
\tilde{\alpha} = \frac{(1 + n^{2})^{2}}{2},
\quad
\tilde{\beta} = \tilde{\delta} = 0,
\quad
\tilde{\gamma} = -\frac{1}{2}.
\eqno(5.11)
$$
Based on reference$^{21}$ and using the result of
Proposition 6, we can find in our case the relation between
equations P3 with $\gamma \delta \neq0$
$$
w'' = \frac{w^{'2}}{w} - \frac{w'}{w} +
\frac{1}{z}( \alpha \gamma w^{2} + \beta) +
\gamma^{2} w^{3} + \frac{\delta}{w},
\eqno(5.12)
$$
and P5 with coefficients given by (5.11).
Indeed, the third Painlev\'e equation (5.12) can be written as
a first
order system of ODEs
$$
\begin{array}{c}
z w' = (\epsilon \alpha  -1) w + \epsilon \gamma z w^{2} + z
v,
\quad \epsilon = \pm 1,   \\
   \\
zw v' = \beta w + \delta z +(\epsilon \alpha  -2) w v + z
v^{2}.
\end{array}
\eqno(5.13)
$$
>From system (5.13), the elimination of $w$ gives
$$
v'' - \frac{v}{v^{2} + \delta} v^{'2} + \frac{v'}{z}
+ \frac{\beta^{2} - (2- \epsilon \alpha )^{2} \delta}{z^{2}
(v^{2} + \delta)} v + \epsilon \gamma (v^{2} + \delta)
- \frac{2 \delta \beta}{z^{2}} \frac{(\epsilon \alpha
-2)}{v^{2} + \delta}
+ \frac{\beta}{z^{2}}(\epsilon \alpha - 2) =0.
\eqno(5.14)
$$
By a homographic transformation of the dependent variable
and a change of the independent variable
$$
v =-i \sqrt{\delta} \frac{y+1}{y-1},
\quad
z = \sqrt{2 \tau},
\eqno(5.15)
$$
we obtain from (5.14) equation P5
$$
y'' + \frac{3y-1}{2y(y-1)} y^{'2} + \frac{y'}{\tau}
+ \frac{1}{32 \delta \tau^{2}} [ (y^{2} -1) (A y
+ \frac{B}{y})] + \frac{\epsilon}{\tau} \gamma (-\delta)^{1/2}
y =0,
\eqno(5.16)
$$
where $A$ and $B$ are defined as follows
$$
\begin{array}{c}
A= \beta^{2} + 4 (-\delta)^{1/2} \beta - \delta \alpha^{2} -4
\delta
- 2 (-\delta)^{1/2} \epsilon \alpha \beta + 4 \epsilon \delta
\alpha, \\
   \\
B = \delta \alpha^{2} -2 (-\delta)^{1/2} \epsilon \alpha \beta
+ 4 (-\delta)^{1/2} \beta +4 \delta - \beta^{2} - 4 \epsilon
\delta \alpha.  \\
\end{array}
\eqno(5.17)
$$

\par
{\bf Proposition 5.} Let $y= y(z)$ be a solution of the fifth
Painlev\'{e} equation (5.16) with parameters given by (5.17),
such that
the function
$$
r(z) = w' -(\epsilon \alpha -1) \frac{w}{z} - \epsilon \gamma
w^{2} -1 \neq 0,
$$
does not vanish. Then the function
$$
S( \tau) = 1 - 2 r^{-1} (\sqrt{2 \tau}),
\eqno(5.18)
$$
is a solution of the third Painlev\'{e} equation  (5.12) with
paramters $\gamma \neq 0$ and $\delta =-2$.

The $\tau$-functions for the rational class of solutions
of the Painlev\'{e} equation P3 can be constructed$^{22,23}$
in terms of the Umemura polynomials $T_{n} = T_{n} (z,l)$
which are determined by a sequence of polynomials in $z$
and defined through the recurrence relation
$$
T_{n+1} T_{n-1} =
(\frac{z}{8} -l + \frac{3}{4} n) T_{n}^{2} +
\frac{\partial T_{n}}{\partial z} T_{n} +
z [ \frac{\partial^{2} T_{n}}{\partial z^{2}} T_{n}
- (\frac{\partial T_{n}}{\partial z})^{2}],
\eqno(5.19)
$$
with initial conditions $T_{0} = T_{1} =1$.
Based on reference $^{22}$, we have

{\bf Proposition 6.} For the existence of rational solutions
of
equation P3 of the form
$$
w(z) = \frac{T_{n+1}(z, l-1) T_{n} (z,l)}
{T_{n+1} (z,l) T_{n} (z, l-1)},
\eqno(5.20)
$$
where the Umemura polynomials $T_{n} = T_{n} (z,l)$
satisfy recurrence relation (5.19),
it is necessary and sufficient that the parameters of
equation P5 satisfy
$$
\alpha = 4(n + l), \quad
\beta = 4 (n-l),   \quad
\gamma= -\delta = 4.
$$

Note that from system (5.13) and the transformation (5.15),
there is a connection between solutions $w$ of the third
Painlev\'e equation (5.12) and the solutions $y$ of the
fifth Painlev\'e equation (5.16),
$$
y = \frac{z w' - [ (4 \epsilon (n+l) -1) + 4 \epsilon z w]w +
2z}
{z w' - [ (4 \epsilon ( n+l) -1) + 4 \epsilon z w]w -2z}.
\eqno(5.21)
$$
Substituting the rational solutions (5.20) into
formula (5.21) and next replacing the $w$ which appears in
(5.3)
by the function $u$ so obtained, we get
multivortex solutions of equations (2.6).
Consequently, by applying Proposition 2 to the so
obtained multivortex solution of (2.6),
certain classes of solutions of system (2.1)
can be found.

Another class of vortex solutions to CSG equations
(2.5) can be provided if we define functions $u$ and $v$
in the polar form following
$$
u = A_{n} (r) e^{in \theta},  \quad
v= A_{n-1}(r)  e^{i(n-1) \theta},
\quad  n \in \mathbb Z.
\eqno(5.22)
$$
which transforms the CSG system (2.5) into
$$
\begin{array}{c}
(i) \quad \dss \frac{d A_{n}}{dr} + \dss \frac{n}{r} A_{n}
= (1 - A_{n}^{2}) A_{n-1},     \\
   \\
(ii) \quad - \dss \frac{d A_{n-1}}{dr} + \dss \frac{(n-1)}{r}
A_{n-1}
= (1 - A_{n-1}^{2}) A_{n}.
\end{array}
\eqno(5.23)
$$

\par
When $n=1$, the second equation (5.23) is
solved by taking $A_{0} = \pm 1$ and then the first equation
(5.23) becomes a Riccati equation which
can be linearized by a Cole-Hopf transformation
and solved in terms of Bessel
functions. The vortex solution (5.22)
takes the form
$$
u = \frac{I_{1}(r)}{I_{0} (r)} e^{ i \theta},
\quad
v = \epsilon = \pm 1,
\eqno(5.24)
$$
where $I_{1}$ is the Bessel function of the first order, that
is,
$I_{1}=I_{0}' (r)$, and the prime denotes differentiation
with respect to $r$. Such a reduction of (5.23) has been
recently obtained$^{19}$.
Consequently, from transformation (2.4), we get
$$
\psi_{1} = \frac{I_{1} (r)}{I_{0} (r)} e^{i \theta}
\bar{\psi}_{2},
\quad \varphi_{1} = \epsilon \bar{\varphi}_{2}.
\eqno(5.25)
$$
Substituting (5.25) into Weierstrass system (2.1) and solving
the
resulting equations, we obtain
$$
\varphi_{2} = F(r e^{-i \theta}),
\eqno(5.26)
$$
where $F$ is an arbitrary function of one variable $r e^{-i
\theta}$
and the function $\psi_{2}$ satisfies the PDE
$$
\frac{\partial \psi_{2}}{\partial r} + \frac{i}{r}
\frac{\partial \psi_{2}}{\partial \theta} =
2 e^{-i \theta} R(r) |\psi_{2}|^{2} \psi_{2},
\qquad
R(r) = 1- \frac{I_{0}^{'2}(r)}{I_{0}^{2}(r)}.
\eqno(5.27)
$$
Equation (5.27) has a solution of the form
$$
\psi_{2} = g ( v(r) e^{-i \theta}),
\eqno(5.28)
$$
where the functions $g$ and $v$ satisfy the following ODEs,
$$
v' + \frac{v}{r} - R(r) \lambda=0, \qquad
 \dot{g} \lambda + |g|^{2} \, g=0,  \quad
\lambda \in \mathbb C,
$$
and $g$ denotes the derivative of $g$ with respect to
$s = v(r) e^{-i \theta}$.
These two equations can be integrated to give
the following expressions
$$
 v(r) = \frac{\lambda}{r} \int^{r}_{0} \tau R(\tau) \, d \tau,
\qquad
\frac{g^{\lambda}}{\bar{g}^{\bar{\lambda}}} = c.
\eqno(5.29)
$$
>From equations (5.25), (5.26) and (5.28),
we can summarize the results as follows,
$$
\begin{array}{cccc}
\dss \psi_{1} = \frac{I_{0}'(r)}{I_{0} (r)} e^{i \theta}
\bar{g}(\frac{\bar{\lambda} e^{i \theta}}{r} \int_{0}^{r} \tau
R(\tau) \, d \tau),
&  & &
\psi_{2} = g( \dss \frac{\lambda e^{-i \theta}}{r}
\int^{r}_{0}
\tau \, R(\tau) \, d \tau \, ).  \\
   &   &   &      \\
\varphi_{1} = \epsilon F( r e^{-i \theta}),  &  &  &
\varphi_{2} = F(r e^{-i \theta}),
\end{array}
\eqno(5.30)
$$
where $g$ is a function of one variable,
which is restricted by relation (5.29).
Note that the solution for the function $u$ in (5.24)
has the form of a scalar field which has appeared in the
study of the vortex solutions of superconductivity
with asymptotic behavior of the radial part of the
solution going to zero as $r$ goes to zero and
constant as $r$ goes to infinity.
Consequently, solutions (5.30) of the Weierstrass
system (2.1) possess similar asymptotic behavior
when functions $F$ and $g$ are bounded.

{\bf VI. EXAMPLES OF SOLVING THE WEIERSTRASS SYSTEM VIA
THE COMPLEX SINE-GORDON EQUATION}

Now we investigate the possibility of generating
new multi-soliton solutions by taking products of known
solutions of the CSG system (2.6). Thus we can formulate the
following:

{\bf Proposition 7.} Suppose $u$ is a solution of equation
(2.6) with constant modulus $|u|^{2} = |c|^{2} \neq 1$.
Suppose also that a complex function $w$ exists which
satisfies
$|w|^{2} =1$, and the differential constraint equation
$$
 u (\partial \bar{\partial} w \mp \frac{\bar{w} |c|^{2}}
{1 \pm |c|^{2}} (\partial w)(\bar{\partial} w))
+ \frac{1}{1 \pm |c|^{2}} (\bar{\partial} u
\partial w + \partial u \bar{\partial} w) = 0.
\eqno(6.1)
$$
Then the product function $U= u \, w$ is a solution of system
(2.6).

{\bf Proof:} Differentiating the function $U=uw$, we get the
following expressions
$$
\bar{\partial} (uw) = (\bar{\partial} u) w + u (\bar{\partial}
w),
\quad
\partial (uw) = (\partial u) w +(\partial w) u,
$$
$$
\partial \bar{\partial} (uw) = (\partial \bar{\partial} u) w
+ (\bar{\partial} u)(\partial w)
+ (\partial u) (\bar{\partial} w) + u (\partial \bar{\partial}
w).
$$
Substituting $U$ into equation (2.6),
using $|u|^{2} = |c|^{2}$ and $|w|^{2}=1$, we obtain that
$$
(\partial \bar{\partial} u) w + (\bar{\partial} u)(\partial w)
+ (\partial u) (\bar{\partial} w) + u (\partial \bar{\partial}
w)
$$
$$
\mp \frac{\bar{u} \bar{w}}{1 \pm |c|^{2}} ((\bar{\partial} u)
w +
u (\bar{\partial} w))((\partial u)w + u (\partial w))
+ \frac{u w}{4} ( 1 \pm |c|^{2})
\eqno(6.2)
$$
$$
= (\partial \bar{\partial} u)w + (\bar{\partial} u)
(\partial w) + (\partial u)(\bar{\partial} w) + u(\partial
\bar{\partial} w)
$$
$$
\mp \frac{\bar{u}w}{1 \pm |c|^{2}} (\partial u)(\bar{\partial}
u)
\mp \frac{|c|^{2}}{1 \pm |c|^{2}}(\partial u)(\bar{\partial}
w)
\mp \frac{|c|^{2}}{1 \pm |c|^{2}} (\bar{\partial} u)(\partial
w)
\mp \frac{|c|^{2} u \bar{w}}{1 \pm |c|^{2}}(\partial
w)(\bar{\partial} w)
+ \frac{uw}{4} (1 \pm |c|^{2}).
$$
Substituting the second derivative $\partial \bar{\partial} u$
from
equation (2.6) into equation (6.2) and next
collecting terms with respect to first derivatives
of $u$ and $w$ and simplifying, we obtain the
differential constraint (6.1). QED

In the case of the CShG equation, the constant $c$
need not necessarily have modulus different from one,
since there is no singularity in the $(1+|c|^2)^{-1}$
term in this case.

At this point, we would like to illustrate Proposition 10 for
constructing a solution to system (2.6) with an elementary
example. The simplest solution of analytic type, a vacuum
solution,
is given by
$$
u = c e^{(\bar{A} z - A \bar{z})},
\eqno(6.3)
$$
where $c$ and $A$ are complex constants.
By substituting the solution $u$ into (2.6), it is easy to
show that this
is a solution provided that the following constraint holds
between $c$ and $A$,
$$
2 \epsilon |A| = 1 \pm |c|^{2}, \qquad  \epsilon = \pm 1.
$$
Suppose that $f$ is a complex valued function of one
complex variable $z$ and define the function $w$ as follows,
$$
w= \frac{f(z)}{\bar{f} (\bar{z})},
$$
such that $f(z)$ satisfies the constraint (6.1), namely,
$$
\partial f(z) \bar{\partial} \bar{f} (\bar{z})
+ A \partial f(z) \bar{f} (\bar{z})
+ \bar{A} f(z) \bar{\partial} \bar{f} ( \bar{z}) = 0.
$$
Then Proposition 7 implies that the function
$$
U = c e^{(\bar{A} z - A \bar{z})}
\frac{f(z)}{\bar{f} (\bar{z})},
$$
is also a solution to system (2.6) and represents
a one-soliton solution.

Let us illustrate our considerations by an example.
of the exponential solution (6.3) for $A=-ia$.
We show that we can use this to obtain a solution to
system (2.1). First of all, since eliminating $u$ from the
pair of equations in (2.5) results in an equation which is
identical to (2.6) but with $u$ replaced by $v$, we can
assign the solution obtained from the second order equation
to either the $u$ or the $v$ variable. Let us take $u=c
e^{ia(a +\bar{z})}$.
The second function $v$ is obtained from (2.5), that is
$$
u = c e^{i a ( z + \bar{z})},
\qquad
v = \frac{2 \partial u}{1 \pm |u|^{2}} = i c e^{i a (z +
\bar{z})}.
\eqno(6.4)
$$
These satisfy (2.5) provided that $2a = 1 \pm |c|^{2}$. From
(2.4)
we can write $\psi_{1} = u \bar{\psi}_{2}$ and
$\varphi_{1} = v \bar{\varphi}_{2}$,
and calculate the quantities $Q_{1,2}$ from (2.1)
$$
Q_{1} = |\psi_{2}|^{2} ( 1 \pm |c|^{2}), \quad
Q_{2} = |\varphi_{2}|^{2} ( 1 \pm |c|^{2}).
\eqno(6.5)
$$
Now, we will show
that we can find a particular class of solutions which satisfy
(2.1).
First from equations (2.1) with (6.5), the equation
$$
\bar{\partial} \psi_{2} = |\psi_{2}|^{2} (1 \pm |c|^{2})
$$
is satisfied by a function of the form $\psi_{2}= e^{b(z-
\bar{z})}$
provided that $b=-2a$ to be consistent with the condition
$2a = 1 \pm |c|^{2}$. Similarly, a function of the form
$\varphi_{2}
= e^{2a (z - \bar{z})}$ satisfies the equations $\partial
\varphi_{2}
= Q_{2} \varphi_{2}$. The equations for $\psi_{1}$ and
$\varphi_{1}$
in (2.1) can also be
integrated, and the set of functions below
$$
\begin{array}{cc}
\psi_{2} = c e^{ia (z +\bar{z})} e^{2a(z- \bar{z})}, &
\psi_{2} = e^{-2a (z - \bar{z})},    \\
     &      \\
\varphi_{1} = ic e^{ia(z+ \bar{z})} e^{-2a(z- \bar{z})},  &
\varphi_{2} = e^{2a(z- \bar{z})}
\end{array}
\eqno(6.6)
$$
constitutes a solution to system (2.1).
Substituting (6.6) into (2.3) and integrating,
we obtain the parametric form of a
surface. One obtains that
$$
X^{1} = -a(z - \bar{z})^2 + 2 a |z|^{2} = ay^{2} + 2 a r^{2},
\quad
X^{2} =  a(z - \bar{z})^2 - 2a |z|^{2} = -a y^{2} - 2a r^{2},
$$
$$
X^{3} = 0,
\quad
X^{4} = i |c|^{2} (z- \bar{z}) = - |c|^{2} y.
\eqno(6.7)
$$
where we set $z - \bar{z} = iy$ and $|z|^{2}= r^{2}$.
Treating $y$ and $r>0$ as parameters, one can plot
$X^{1}$, $X^{2}$ and $X^{4}$ to give the surface given in
Figure 1,
which has the form of a parabolic cylinder.
Moreover, from (6.12), we can calculate the
components of the induced metric
$$
g_{zz} = \sum_{i=1}^{4} (X^{i}_{z})^{2}
= 8 a^2 (z^2 -4 |z|^{2} + 4 \bar{z}^{2}) - |c|^{4} =
\bar{g}_{\bar{z}
\bar{z}},
$$
$$
g_{z \bar{z}} = \sum_{i=1}^{4} X^{i}_{z} X^{i}_{\bar{z}}
= 8 a^{2} (2 z + \bar{z})(z - 2 \bar{z}) + |c|^{4}.
$$

A procedure for obtaining solutions to (2.1) as a result of
using
Proposition 10 can be developed from the above example. The
new
product solution can be called either $u$ or $v$. One
substitutes this
in the corresponding equation in (2.5)
to obtain the remaining unknown solution $v$ or $u$. Using
these results
in (2.4) to eliminate $(\bar{\psi}_{2}, \bar{\varphi}_{2})$,
one tries to integrate the
nonlinear system (2.1) to obtain the
required complex functions $\psi_{1}$ and
$\varphi_{1}$, which are solutions of the Weierstrass system
(2.1).

{\bf IV. FINAL REMARKS.}

Equations (2.1) and (2.5) under investigation here have
long been of interest in field theory.
In particular, there has been the extensive use of applying
soliton solutions to construct
models of extended particles $^{6,24}$. The Sine-Gordon
equation,
it seems is the only Lorentz-invariant, nonlinear
equation whose initial value problem has been solved $^{4}$.
This equation also describes a completely integrable
Hamiltonian system. It would certainly be of great
interest to find other Lorentz invariant integrable
systems. Of more recent interest is the study of
vortex tubes $^{25}$. The motion of vortex tubes in an
inviscid incompressible fluid is described by the Biot-Savard
law.
The recently proposed localized induction equation is the
simplest model to capture the leading order behavior
of the three-dimensional self-induced motion of a
vortex filament. This type of equation is in fact
related to the cubic nonlinear Schr\"{o}dinger equation
for a complex variable, and implies that the localized
induction equation is completely integrable.
Note that from solutions (3.36), when the functions
$F$ and $g$ are real polynomials in a single variable,
the vortex structure of the solutions is preserved at the
level of the functions $\psi_{\alpha}$ and
$\varphi_{\beta}$. These functions are applied to
generate surfaces. Consequently,
by plotting such results in two or three dimensions,
these surfaces could model a vortex filament in such
a fluid $^{24,25}$.

We have presented a new approach to the study of the
Weierstrass system (2.1) in connection with
CShG and CSG equations (2.6).
It proved to be particularly
effective in constructing multivortex solutions of (2.1)
in terms of $\tau$-functions based on rational solutions
of the third and fifth Painlev\'{e} equations.
It is worth noting that the approach to the
Weierstrass system (2.1) proposed here can be applied with
some necessary modifications, to more general cases
of Weierstrass type systems describing more diverse surfaces
immersed in multi-dimensional Minkowski and pseudo-Riemannian
spaces.
The task of obtaining new types of minimal
surfaces described by system (2.1)
will be undertaken in our future work.

{\bf ACKNOWLEDGMENTS:}

This work was supported in part by
research grants from NSERC of Canada and Fonds FCAR du
Gouvernement du Qu\'{e}bec.

{\bf REFERENCES.}

\noindent
$^{1}$ K. Pohlmeyer, Integrable Hamiltonian systems and
interactions
through quadratic constraints, Commun. Math. Phys. {\bf 46},
207-221, (1976). A. Neveu and N. Papanicolaou,
Integrability of the Classical $[\bar{\psi}_{i}
\psi_{i}]^{2}_{2}$
and $[\bar{\psi}_{i} \psi_{i}]^{2}_{2}- [\bar{\psi}_{i}
\gamma_{5} \psi_{i}]_{2}^{2}$ Interactions,
Commun. Math. Phys. {\bf 58}, 31-64, (1978).     \\
$^{2}$ A. I. Bobenko and U. Eitner, Painlev\'e Equations in
the
Differential Geometry of Surfaces, Lecture Notes in
Mathematics, 1753,
Springer Verlag, (2000).  \\
$^{3}$ V. G. Makhankov and O. K. Pashaev, ``Integrable
Pseudospin
Models in Condensed Matter'', Sov. Sci. Rev. Math. Phys.
{\bf 9}, 1 (1992).  \\
$^{4}$ F. Lund, T. Regge, Unified approach to string
and vortices with soliton solutions, Phys. Rev. {\bf D15},
1524-1534, (1977).
F. Lund, Example of a Relativistic, Completely Integrable,
Hamiltonian System,
Phys. Rev. Letts. {\bf 38}, 1175, (1977).
F. Lund, Classically Solvable Field Theory Model,
Ann. of Phys., 251, (1978).    \\
$^{5}$ B. S. Getmanov, Integrable Model of a Nonlinear Complex
Scalar
Field with Nontrivial Asymptotic Behavior of Soliton
Solutions,
Sov. Phys., JETP Lett. {\bf 25}, 119, (1977).      \\
$^{6}$ W. Zakrzewski, Low Dimensional Sigma-Models (Hilger,
New York,
1989).   \\
$^{7}$ V. L. Kosevich and L. P. Pitaevskii, Sov. Phys. JETP,
{\bf 34},
858, (1958). E, M. Lifshitz and L. P. Pitaevskii,
Statistical Physics, Part 2 (Pergamon Press, Oxford, 1980).
\\
$^{8}$ D. Nelson, T. Piran, and S. Weinberg, Statistical
Mechanics
of Membranes and Surfaces (World Scientific, Singapore, 1992).
\\
$^{9}$ Ou-Yang Zhong-Can, Liu Ji-Xing and Xie Yu-Zhang,
Geometric Methods in the Elastic Theory of Membranes in Liquid
Crystal Phases, World Scientific, (1999).   \\
$^{10}$ G. Darboux, Lecons sur la th\'eorie g\'en\'erale des
surfaces,
vol 3, Lignes g\'eod\'esiques et courbure g\'eod\'esique
param\`etres diff\'erentiels-deformation des surfaces,
Ed. Gauthier-Villars, Paris, 1894.   \\
$^{11}$ B. Konopelchenko and G. Landolfi, Generalized
Weierstrass
representation for surfaces in multi-dimensional Riemann
spaces,
J. Geom. Phys., {\bf 29}, 319-333, (1999).    \\
$^{12}$ J. Tafel, Two-dimensional reductions of the self-dual
Yang-Mills equations in self-dual spaces, J. Math. Phys.
{\bf 34}, 5,(1993), 1892-1907.  \\
$^{13}$ I. Barashenkov, B. Getmanov, The unified approach to
integrable relativistic equations: soliton solutions over
nonvanishing backgrounds, J. Math. Phys. {\bf 34,7} (1993),
3054-3072.   \\
$^{14}$ M. J. Ablovitz, A. Ramani and H. Segur, J. Math. Phys.
{\bf 21}, 715, (1980).   \\
$^{15}$ J. Weiss, M. Tabor and G. Carnevale, J. Math. Phys.,
{\bf 24}, 522, (1983), J. Weiss, J. Math. Phys., {\bf 25},
2226, (1984).   \\
$^{16}$ M. Musette, Painlev\'{e} analysis for nonlinear
partial differential equations, chapter 8, 517-572 in The
Painlev\'{e} Property One Century Later, Ed. R. Conte,
Springer Verlag, New York, 1999.   \\
$^{17}$ M. A. Weiss, Dissertation, Flots isospectraux et
moments dans
des alg\`ebras de lacets appliqu\'es \`a des \'equations
diff\'erentielles
nonlin\'eaires, CRN Preprint, Universit\'e de Montr\'eal,
1993. \\
$^{18}$ M. Bruschi and O. Ragnisco, Nonlinear evolution
equations
associated with the chiral-field spectral problem,
Nuovo-Cimento-B (11),
{\bf 88}, 119-139, (1985).   \\
$^{19}$ I. V. Barashenkov and D. E. Pelinovsky, Exact Vortex
Solutions
of the Complex Sine-Gordon Theory on the Plane, Phys. Letts.
{\bf 436},
117-124, (1998).   \\
$^{20}$ R. Conte, The Painlev\'{e} approach to nonlinear
ordinary differential equations, chapter 3, 77-180 in The
Painlev\'{e} Property One Century Later, Ed. R. Conte,
Springer Verlag, New York, 1999.   \\
$^{21}$ V. Gromak, B\"{a}cklund transformations of
Painlev\'{e}
equations and their applications, chapter 12, 687-734,
The Painlev\'{e} Property, One Century Later Ed. R. Conte,
Springer Verlag, New York, 1999.    \\
$^{22}$ K. Kajiwara, T. Masuda, On the Umemura polynomials
for the Painlev\'{e} III equation, Phys. Letts. {\bf A 260},
(1999), 462-467.  \\
$^{23}$ K. Okamoto, The Hamiltonians associated to the
Painlev\'{e}
equations, chapter 13, 735-789, The Painlev\'{e} Property
One Century Later, Ed. R. Conte, Springer Verlag,
New York, 1999.   \\
$^{24}$ R. Rajamaran, Solitons and Instantons (North-Holland,
Amsterdam, 1982).   \\
$^{25}$ Y. Fukumoto and M. Miyajima, The Localized Induction
Hierarchy and the Lund-Regge Equation, J. Phys. A: Math. Gen.
{\bf 29}, (1996), 8025.  \\

\newpage
\begin{center}
{\bf Figure Captions.}
\end{center}

Figure 1. The surface which corresponds to the solution (6.6)
of the CShG equation
and coordinates (6.7) for the choice of constants $a=1$ and
$c= 1$.

\newpage
\includegraphics{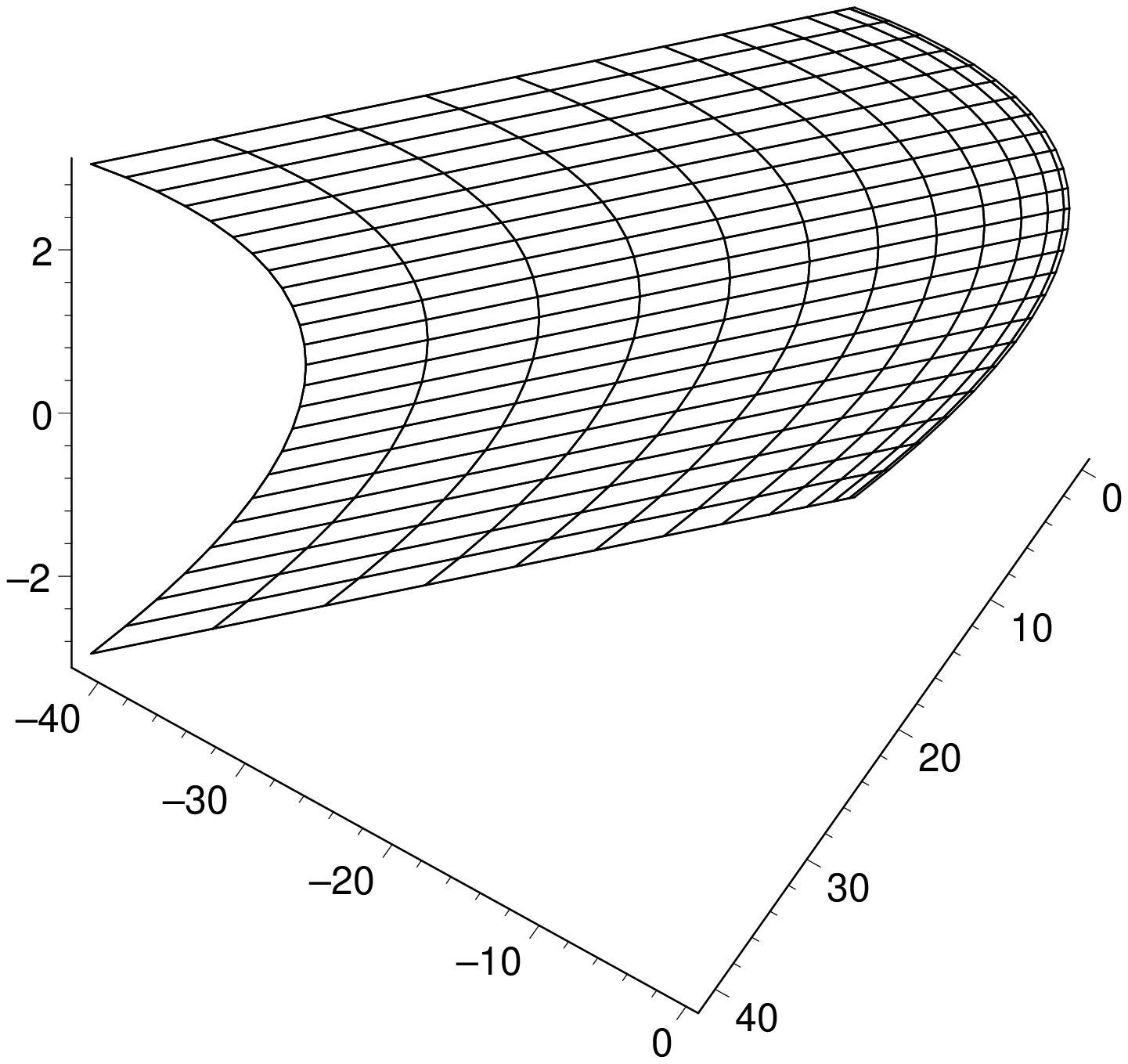}
\end{document}